\documentclass{PoS}

\title{Determination of latent heat at the finite temperature phase transition of SU(3) gauge theory}

\ShortTitle{Determination of latent heat at the finite temperature phase transition of SU(3) gauge theory}

\author{\speaker{Shinji Ejiri}%
\\
Department of Physics, Niigata University, Niigata 950-2181, Japan\\
E-mail: \email{ejiri@muse.sc.niigata-u.ac.jp}}
\author{Ryo Iwami, Mizuki Shirogane and Naoki Wakabayashi\\
Graduate School of Science and Technology, Niigata University, Niigata 950-2181, Japan
}

\author{Kazuyuki Kanaya
\\
Center for Integrated Research in Fundamental Science and
Engineering (CiRfSE), University of Tsukuba, Tsukuba, Ibaraki 305-8571, Japan
}
\author{Masakiyo Kitazawa\\
Department of Physics, Osaka University, Osaka 560-0043, Japan\\
J-PARC Branch, KEK Theory Center, Institute of Particle and Nuclear Studies,
KEK, 203-1, Shirakata, Tokai, Ibaraki, 319-1106, Japan
}
\author{Hiroshi Suzuki\\
Department of Physics, Kyushu University, 744 Motooka, Fukuoka 819-0395, Japan
}
\author{Yusuke Taniguchi\\
Center for Computational Sciences, University of Tsukuba, Tsukuba, Ibaraki 305-8571, Japan
}
\author{Takashi Umeda\\
Graduate School of Education, Hiroshima University, Higashihiroshima, Hiroshima 739-8524, Japan
}

\abstract{
We calculate the energy gap (latent heat) and pressure gap between the hot and cold phases 
of the SU(3) gauge theory at the first order deconfining phase transition point.
We perform simulations around the phase transition point with the lattice size in the temporal direction $N_t=6,$ $8$ and $12$ and extrapolate the results to the continuum limit. 
The energy density and pressure are evaluated by the derivative method with nonperturabative anisotropy coefficients. 
We find that the pressure gap vanishes at all values of $N_t$.
The spatial volume dependence in the latent heat is found to be small on large lattices. 
Performing extrapolation to the continuum limit, we obtain 
$\Delta \epsilon/T^4 = 0.75 \pm 0.17$
and
$\Delta (\epsilon -3 p)/T^4 = 0.623 \pm 0.056.$
We also tested a method using the Yang-Mills gradient flow. 
The preliminary results are consistent with those by the derivative method within the error.
}

\FullConference{34th annual International Symposium on Lattice Field Theory\\
         24-30 July 2016\\
         University of Southampton, UK}

\begin{document}

\section{Introduction}
\label{sec:intro}

We study thermodynamic properties near the first order phase transition in the finite temperature 
SU(3) gauge theory (the quenched approximation of QCD).
First order phase transitions are expected in interesting systems such as the high density 
region of QCD and the many-flavor QCD aiming at construction of a walking technicolor model.
The SU(3) gauge theory at finite temperature has a first order transition and is a good testing ground to develop techniques to investigate thermodynamic quantities around the phase transition.

At a first order phase transition point, two phases coexist at the same time.
To keep a balance between them, the pressure must be the same in the two phases.
On the other hand, the energy density is different in these phases.
The difference is the latent heat which is one of the most important physical quantities characterizing the first order phase transition.
In numerical studies of QCD thermodynamics, the integral method is widely used. 
However, the pressure gap is set to be zero in the integral method.
Because we consider that confirmation of the vanishing pressure gap is important 
to develop the technique to calculate the thermodynamic quantities, 
we adopt the derivative method and the gradient flow method proposed by Ref.~\cite{suzuki13} 
in this study. 

In the derivative method, the derivatives of gauge coupling constants with respect to 
the anisotropic lattice spacings, 
which we call the anisotropy coefficients, are required \cite{karsch}.
Since the perturbative coefficients are known to lead to
pathological results such as negative pressure, 
we calculate the anisotropy coefficients nonperturbatively following Ref.~\cite{ejiri98}. 
Simulations are performed on lattices with the temporal extension $N_t =6,$ $8$ and 12 in Sec.~\ref{sec:results}. 
We carry out the continuum extrapolation of the latent heat. 
We also investigate the spatial volume dependence.

On the other hand, new method to compute thermodynamic quantities is proposed on the basis of the Yang-Mills gradient flow \cite{suzuki13}. 
Using the gradient flow method, we calculate the latent heat and pressure gap at the phase transition point in Sec.~\ref{sec:flow}.
Because the latent heat is one of the most well-defined quantities in the SU(3) gauge theory and the pressure gap must vanish, 
the calculation of the latent heat is a very good test to confirm the reliability of the gradient flow method.

%In the next section, we explain the formulation of the derivative method.
%The numerical results of the latent heat and pressure gap are shown in Sec.~\ref{sec:results}.
%The preliminary results by the gradient flow method are presented in Sec.~\ref{sec:flow}.
%The section \ref{sec:} is the conclusion.

\section{Derivative method for the calculation of the latent heat and the pressure gap}
\label{sec:derivative}

The energy density $\epsilon$ and the pressure $p$
are defined by the derivatives of the partition function $Z$ 
in terms of the temperature $T$ 
and the physical volume $V$ of the system
\begin{eqnarray}
\epsilon = - \frac{1}{V} \left. \frac{\partial \ln Z}{\partial \,T^{-1}} \right|_{V}, 
\hspace{5mm} 
p       = T \left. \frac{\partial \ln Z}{\partial \, V} \right|_{T}. 
\label{eqn:ep}
\end{eqnarray}
On a lattice with a size $N_s^3\times N_t$, the volume and temperature are given by 
$V = (N_s a_s)^3$ and $T = 1 / (N_t a_t)$, 
with $a_s$ and $a_t$ the lattice spacings in spatial and temporal directions.
Because $N_s$ and $N_t$ are discrete parameters, 
the partial differentiations in Eq.~(\ref{eqn:ep}) are performed 
by varying $a_s$ and $a_t$ independently on anisotropic lattices \cite{karsch}.
The anisotropy on a lattice is realized by introducing 
different coupling parameters in temporal and spatial directions.
For an SU($N_c$) gauge theory, the standard plaquette action 
on an anisotropic lattice is given by
\begin{eqnarray}
 S = -\beta_s \sum_{i<j \ne 4} \sum_{x} P_{ij}(x) 
     -\beta_t \sum_{i \ne 4} \sum_{x} P_{i4}(x),
\end{eqnarray}
where
$ P_{\mu \nu}(x) = N_{c}^{-1} {\rm Re \ Tr} 
[ U_{\mu}(x) U_{\nu}(x+\hat{\mu})
   U^{\dagger}_{\mu}(x+\hat{\nu}) U^{\dagger}_{\nu}(x) ]$
is the plaquette in the $(\mu,\nu)$ plane.
With this action, 
the energy density is given by 
\begin{eqnarray}
\epsilon &=& - \frac{3 N_t^{4} T^{4}}{\xi^3} 
  \left\{ \left(a_t \frac{\partial \beta_s}{\partial a_t} 
 -\xi \frac{\partial \beta_s}{\partial \xi}\right) 
 \left(\langle P_s \rangle - \langle P \rangle_0\right) + 
  \left(a_t \frac{\partial \beta_t}{\partial a_t} 
 -\xi \frac{\partial \beta_t}{\partial \xi}\right) 
 \left(\langle P_t \rangle - \langle P \rangle_0\right) \right\}, \label{enrg} 
%\\
%p &=& \frac{N_t^{4} T^{4}}{\xi^3} 
%  \left\{\xi \frac{\partial \beta_s}{\partial \xi} \,
% \left(\langle P_s \rangle  - \langle P \rangle_0\right)
% +\xi \frac{\partial \beta_t}{\partial \xi} \,
% \left(\langle P_t \rangle - \langle P \rangle_0\right) \right\}, \label{prs}
\end{eqnarray}
where $\langle P_{s(t)} \rangle$ is
the space(time)-like plaquette expectation value %, 
%\begin{eqnarray}
%P_s = \frac{1}{3N_{\rm site}} \sum_{i<j \ne 4} \sum_x P_{ij}(x) 
%\hspace{5mm} {\rm and} \hspace{5mm} 
%P_t = \frac{1}{3N_{\rm site}} \sum_{i \ne 4} \sum_x P_{i4}(x),
%\end{eqnarray}
and $\langle P \rangle_0$ is the plaquette expectation value 
on a zero temperature lattice.
For later convenience, we have chosen $a_t$ and 
$\xi \equiv a_s/a_t$ as 
independent variables to vary the lattice spacings.

The derivatives of the gauge coupling constants
with respect to the anisotropic lattice spacings,
%\begin{eqnarray}
$
a_t \frac{\partial \beta_s}{\partial a_t},
\ 
a_t \frac{\partial \beta_t}{\partial a_t},
\ 
\frac{\partial \beta_s}{\partial \xi},
\ 
\frac{\partial \beta_t}{\partial \xi}
$
%\end{eqnarray}
are called the anisotropy coefficients. 
We need these values of anisotropy coefficients to calculate the energy density and pressure by a simulation. 
On isotropic lattices with $a_s=a_t=a$ and $\xi =1$, the coupling constants satisfy $\beta_s = \beta_t \equiv \beta$
and we have 
%\begin{eqnarray}
$( a_t \frac{\partial \beta_s}{\partial a_t} )_{\xi = 1}
= ( a_t \frac{\partial \beta_t}{\partial a_t} )_{\xi = 1}
= a \frac{d \beta}{d a} 
= 2N_{c} \, a \frac{d g^{-2}}{d a},
$
%\end{eqnarray}
where 
$\beta = 2N_c\,g^{-2}$ and 
$a \frac{d g^{-2}}{d a}$ is 
the beta function at $\xi = 1$,
whose nonperturbative value is well studied by numerical simulations.
Moreover, from the fact that the string tension $\sigma$ is independent of $\xi=a_s/a_t$, 
a combination of the remaining two anisotropy coefficients 
is known to be related to the beta function \cite{karsch} as%
%\footnote{
%In \cite{karsch}, a corresponding equation is given for
%$(\partial \beta_{s (t)} / \partial \xi )_{a_s: {\rm fixed}}$.
%}
\begin{eqnarray}
\left(\frac{\partial \beta_s}{\partial \xi} 
+ \frac{\partial \beta_t}{\partial \xi}\right)
_{a_t : {\rm fixed},\, \xi = 1}  
= \frac{3}{2} \, a \frac{d \beta}{d a}. 
\label{eq:cubicsym}
\end{eqnarray}

The ratio can be determined by measuring the phase transition point 
in the $(\beta_s, \beta_t)$ plane \cite{ejiri98}. 
The transition temperature $T_{c} = 1/[N_t a_t (\beta_s, \beta_t)]$ 
must be independent of the anisotropy of the lattice. 
Therefore, 
when we change the coupling constants, 
$ (\beta_s, \beta_t) \rightarrow 
(\beta_s + d \beta_s, 
\beta_t + d \beta_t) $ on a lattice with fixed $N_t$, 
along the transition curve, % in the $(\beta_s,\beta_t)$ plane,
the lattice spacing in the temporal direction $a_t$ does not change, i.e.
$
d a_{t }= \frac{\partial a_t}{\partial \beta_s} \,
d \beta_s + \frac{\partial a_t}{\partial \beta_t} \,
d \beta_t = 0.
$
Denoting the slope of the transition curve at $\xi = 1$ as $r_t$,
\begin{eqnarray}
r_t =  
\frac{d \beta_s}{d \beta_t} = 
- \left(\frac{\partial a_t}{\partial \beta_t}\right)_{\xi = 1} 
\left/
  \left(\frac{\partial a_t}{\partial \beta_s}\right)_{\xi = 1}
\right. 
=  \left(\frac{\partial \beta_s}{\partial \xi}\right)_{\xi = 1} 
\left/ \left(\frac{\partial \beta_t}{\partial \xi}\right)_{\xi = 1}
\right. .
\label{eq:rtr}
\end{eqnarray}
Therefore, when the value for the beta function is available, 
we can determine these anisotropy coefficients nonperturbatively by measuring $r_t$ 
from the phase transition line in 
the $(\beta_s, \beta_t)$ plane.

The latent heat $\Delta \epsilon$ and pressure gap $\Delta p$,
i.e.  the differences of the energy density and pressure between hot and cold phases, 
can be calculated by performing simulation at the transition temperature with 
$\xi=1$ and separating the configurations into the hot and cold phases.
During a Monte Carlo simulation at the phase transition point, the system flop-flops 
between hot and cold phases. The probability of occurrence of 
mixed states, in which the hot and cold phases coexist in one configuration, 
are small in practice.
We classify the configurations into the hot, cold and mixed phases by the value of  
the order parameter of the confinement, Polyakov loop, measured on each configuration.
Using the slope $r_t$ and the beta function, 
the conventional combinations $\Delta \epsilon-3 \Delta p$ and $\Delta \epsilon + \Delta p$
are given by 
\begin{eqnarray}
\frac{\Delta (\epsilon -3p)}{T^4} &=& 
- 3 N_t^4 \, a \frac{d\beta}{da} \,
\{\langle P_s \rangle_{\rm hot} + \langle P_t \rangle_{\rm hot} 
- \langle P_s \rangle_{\rm cold} + \langle P_t \rangle_{\rm cold} \}, 
\label{eq:e3p} \\
\frac{\Delta (\epsilon + p)}{T^4} &=& 
3 N_t^4 \, a \frac{d\beta}{da} \, 
\frac{r_t-1}{r_t+1} \,
\{ \langle P_s \rangle_{\rm hot} - \langle P_t \rangle_{\rm hot} 
- \langle P_s \rangle_{\rm cold} - \langle P_t \rangle_{\rm cold} \},
\label{eq:emp}
\end{eqnarray}
where $\langle \cdots \rangle_{\rm hot}$ and $\langle \cdots \rangle_{\rm cold}$ mean the expectation values in the hot and cold phases, respectively.
Note that, in the calculations of $\Delta \epsilon$ and $\Delta p$, the zero temperature subtraction is not necessary.

\section{Numerical results by the derivative method}
\label{sec:results}

\paragraph{Numelical simulations}
We perform simulations of the SU(3) gauge theory 
on isotopic lattices $\xi=1$
%i.e. $\beta_s = \beta_t = \beta$, $\xi = a_s/a_t =1$, 
at several $\beta$ points around the deconfining phase transition point. 
The lattice sizes for temporal direction are $N_t=6, 8$ and $12$ with two different volumes for each $N_t$.
The configurations are generated by a pseudo heat bath algorithm followed by 5 over-relaxation sweeps.
The Polyakov loop and the plaquettes are measured every iteration.
Data are taken at 1 to 5 $\beta$ values for each $(N_s,N_t)$ 
and are combined using the multipoint reweighting method \cite{iwami15}.
The details of our simulation parameters are given in Ref.~\cite{shirogane16}.
The statistical errors are estimated by the jack-knife method. 
The bin size is adopted to be 1000, which is much smaller than the typical size of the interval of flip-flops.
The errors are saturated with this bin size.
For continuum and large volume extrapolations, we include the data obtained on 
$36^2 \times 48 \times 6$ lattice by the QCDPAX Collaboration \cite{QCDPAX}.

\begin{figure}[t]
\begin{center}
\vspace*{-8mm}
\centerline{
\includegraphics[width=60mm,clip]{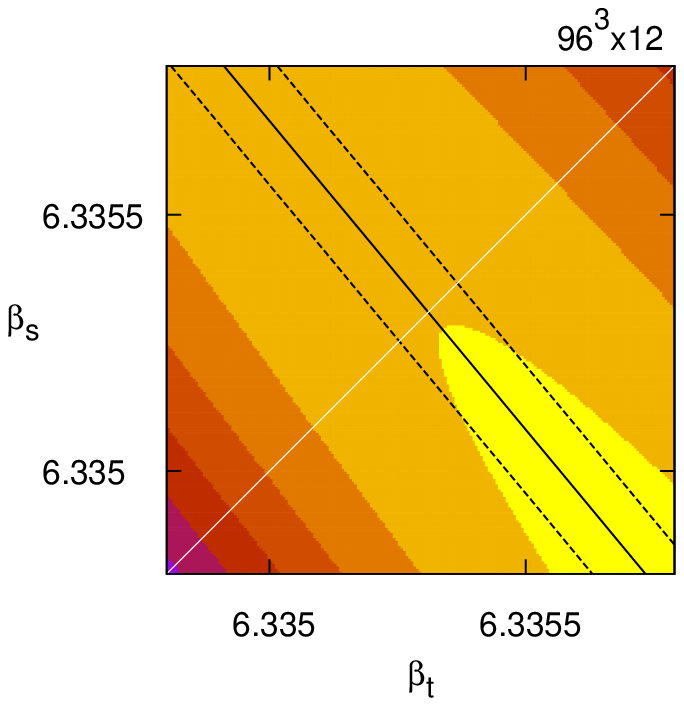}
\hspace*{7mm}
\includegraphics[width=58mm,clip]{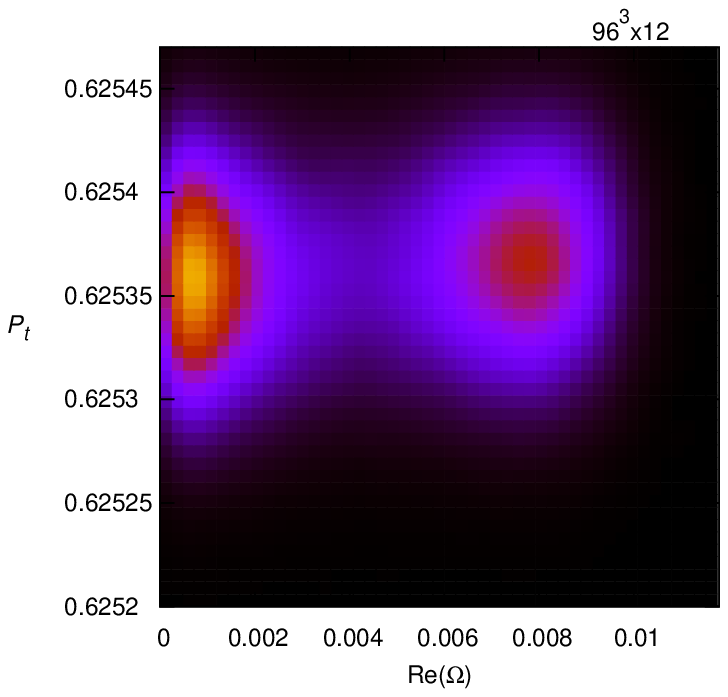}
}
\vspace{-2mm}
\caption{Left: Contour plot of $\chi_{\Omega}$ 
as a function of $(\beta_s, \beta_t)$ obtained on the $96^3 \times 12$ lattice 
\cite{shirogane16}. 
The solid line is the phase transition line and the 
dashed lines are the upper and lower bounds of the error.
Right: Histogram as functions of $(P_t, {\rm Re} \Omega)$ at the transition point 
on the $96^3 \times 12$ lattice \cite{shirogane16}.
%Brighter color means larger probability.
}
\label{fig1}
\end{center}
\end{figure}

\paragraph{Slope of the transition line}
In order to determine the transition line in the coupling 
parameter space $(\beta_s, \beta_t)$, 
we define the transition point as the peak position of the Polyakov loop susceptibility 
%\begin{eqnarray}
$
\chi_{\Omega} = N_s^3\left(\langle \Omega^2 \rangle - \langle \Omega \rangle^2\right) ,
$
%\end{eqnarray}
where $\Omega$ is the rotated Polyakov loop. 
%\begin{eqnarray}
%\Omega = z \, \frac{1}{N_s^3} \sum_{\vec{x}} \frac{1}{N_{c}} 
%{\rm Tr} \prod_{t=1}^{N_t} U_4( \vec{x},t ) ,
%\end{eqnarray}
%$z$ is a $Z(N_c)$ phase factor ($z^{N_c} = 1$) 
The phase facter is rotated such that $\arg(\Omega) \in (-\pi/N_c,\pi/N_c]$.
We compute the Polyakov loop susceptibility $\chi_{\Omega}$ as a function of $(\beta_s, \beta_t)$ 
using the multipoint reweighting method.
The left figure of Fig.~\ref{fig1} is the contour plot of $\chi_{\Omega}$ measured on 
the $64^3 \times 6$ lattice.
Because the transition is of first order for the SU(3) gauge theory, the peak of $\chi_{\Omega}$
is quite clear with our large spatial volumes. 
A brighter color means a larger $\chi_{\Omega}$.
The phase transition point is defined as the peak position of the susceptibility 
for each fixed $\gamma \equiv \sqrt{\beta_t/\beta_s}$.  
The transition line is shown by the solid line in Fig.~\ref{fig1} (left), 
with the dashed lines being their jackknife errors.
We then calculate the slope $r_t$.
The details of the determination of $r_t$ are written in Ref.~\cite{shirogane16}.
We have confirmed that the systematic error caused by the choice of the fit range and the fit function is small.

\paragraph{Beta function}
The beta function $a (d \beta /da)$ is computed from the data of the transition point $\beta_c$ for each $N_t$.
Because the lattice spacing is $a=1/(N_t T_c)$ at $\beta_c$, 
we have $a (d \beta /da) = -N_t (d \beta_c/d N_t)$.
We use our results of $\beta_c(N_t)$ at $N_t=6, 8, 12$
together with the data at $N_t=4$, 10, and $14$--$22$ reported in Ref.~\cite{francis15}.
After performing the extrapolation of $\beta_c$ to the infinite volume limit, 
we fit the data of $\beta_c(N_t,\infty)$ by a polynomial function,
$
\beta_c (N_t,\infty) = \sum_{n=0}^{n_{\rm max}} b_n \, N_t ^{\, n}
$
with $b_n$ being the fit parameters. $n_{\rm max}=5$ is adopted for the final result.
We obtain $a (d \beta /da) = -0.5488(8)$, $-0.6217(8)$ and $-0.7166(26)$ at $\beta_c(N_t,\infty)$ for $N_t=6$, 8 and 12, respectively.

\paragraph{Phase separation at the first order transition}
To evaluate the latent heat and the pressure gap, we need to separate the configurations at the first order transition point into 
the hot and cold phases.
Because $\Delta \epsilon /T^4$ and $\Delta p /T^4$ are proportional to $N_t^4$, 
the gaps in the plaquettes decrease as $1/N_t^4$. 
Thus, a high precision measurement is required at large $N_t$.
In the right panel of Fig.~\ref{fig1}, we show a contour plot of the histogram 
as a function of $(P_t, {\rm Re} \Omega)$ obtained on the $96^3 \times 12$ lattice.
Using the multipoint reweighting method, $\beta$ is adjusted to the transition point. 
The two peaks correspond to the hot and cold phases. 
The peaks are well separated in the ${\rm Re}\Omega$ direction, while they are overlapping in the plaquette directions.

We separate the two phases by introducing cuts in the time history of the Polyakov loop.
To remove short-time-range fluctuations, we average ${\rm Re} \Omega$ over $\pm 250$ configurations around the current configuration number.  
We then identify the hot/cold phase by the value of the time-smeared Polyakov loop.
The configurations in the mixed phase are discarded. 
After the phase separation, we combine the configurations by the multipoint reweighting method to compute the expectation values of the plaquettes in each phase at the transition point.

\begin{figure}[t]
\begin{center}
\vspace*{-5mm}
\centerline{
\includegraphics[width=70mm,clip]{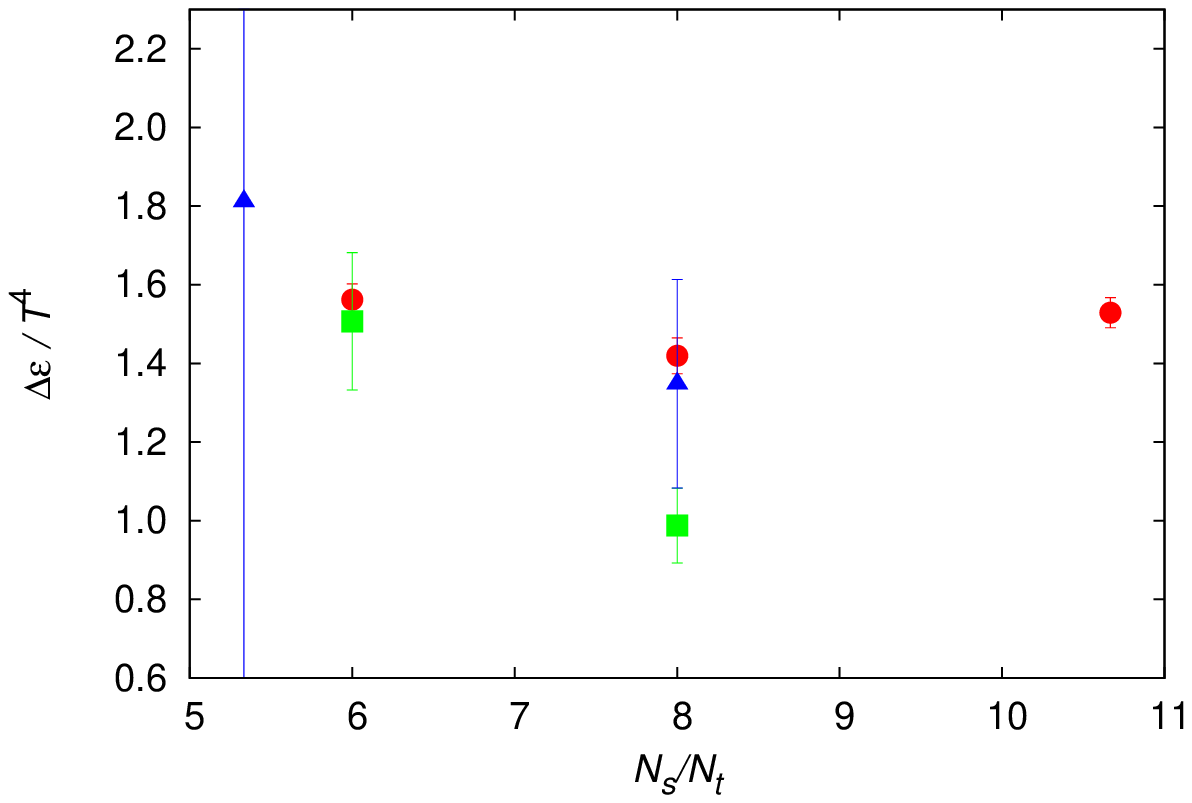}
\hspace*{-3mm}
\includegraphics[width=70mm,clip]{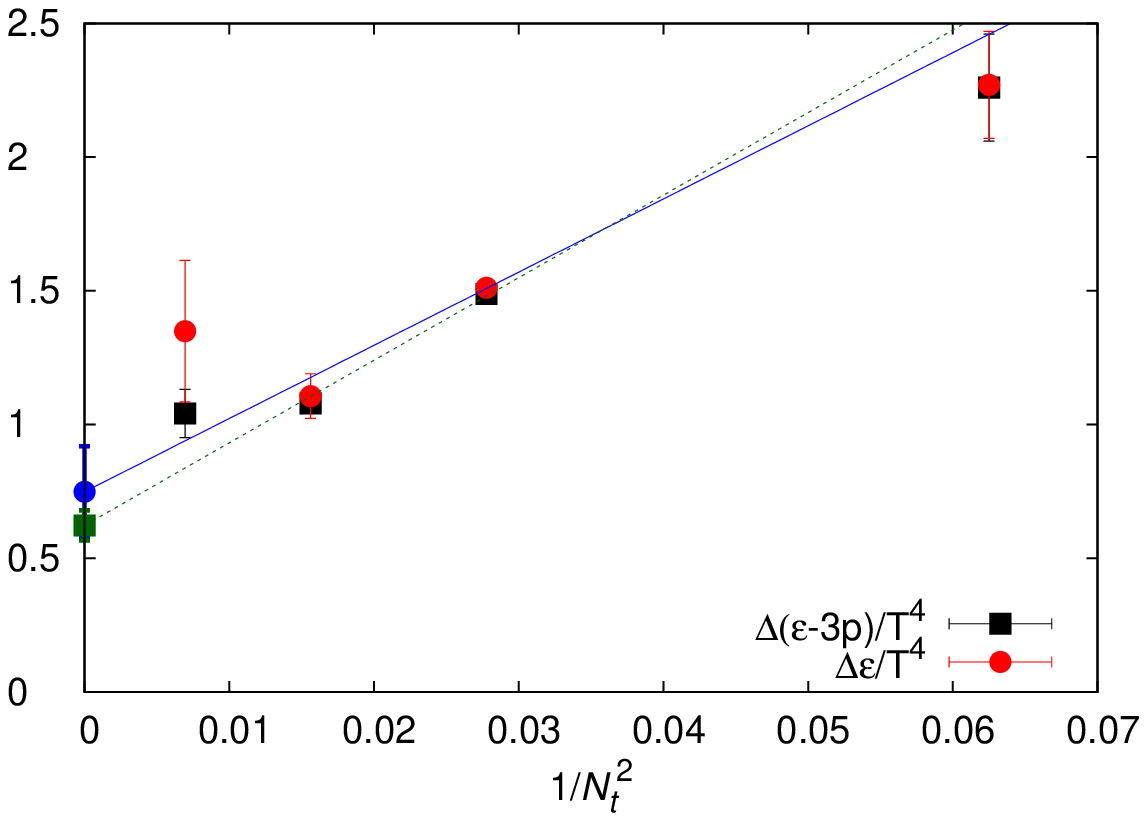}
}
\vspace{-2mm}
\caption{Left: Latent heat $\Delta \epsilon /T^4$ for $N_t=6$ (circle), $8$ (square) and $12$ (triangle) as a function of the aspect ratio $N_s/N_t$ \cite{shirogane16}.
Right: Continuum extrapolation of the latent heat $\Delta \epsilon/T^4$ (circle) and  $\Delta (\epsilon -3 p)/T^4$ (square) using data at $N_t=6$, 8 and 12 \cite{shirogane16}.
The rightmost data at $N_t=4$ are obtained in Ref.~\cite{ejiri98}.
%, with error bars including systematic errors due to the choice of the beta function.
}
\label{fig2}
\end{center}
\end{figure}

\paragraph{Latent heat and pressure gap}
Using the nonperturbative anisotropy coefficients and the plaquette gaps,
we compute the latent heat $\Delta \epsilon$ and the pressure gap $\Delta p$ using Eqs.~(\ref{eq:e3p}) and (\ref{eq:emp}).
In the left figure of Fig.~\ref{fig2}, we plot the results of latent heat $\Delta \epsilon/T^4$ as functions of the spatial volume.
Because the correlation length remains finite at first order transition, 
we expect that $\Delta \epsilon$ at the transition point is independent of the spatial volume on sufficiently large lattices.
The horizontal axis is the aspect ratio $N_s/N_t$, and the results at $N_t=6$, $8$ and $12$ are shown by circle, square and triangle symbols, respectively.
From the results of $N_t=6$ , we find that the latent heat is well stable at $N_s/N_t \ge 6$.
The results at $N_t=8$ and 12 are also consistent with constant, although the errors are large.
We thus perform a constant fit of the data in Fig.~\ref{fig2} (left) at each $N_t$, 
which are plotted by circles in Fig.~\ref{fig2} (right), together with the result of $\Delta (\epsilon -3p)/T^4$ (square). 
Because the anisotropy coefficients are not needed for $\Delta (\epsilon -3p)/T^4$, 
the statistical errors are smaller than those of $\Delta \epsilon/T^4$.

We then extrapolate the results to the continuum limit.
Because the leading lattice artifact in the action is $O(a^2)$ and also the equation of state in the high temperature limit is a function of $N_t^2$, we carry out linear extrapolations in $1/N_t^2$. 
Using the data at $N_t =6$, $8$ and $12$,  we obtain the solid and dashed lines in Fig.~\ref{fig2} (right). The results in the continuum limit are 
$\Delta \epsilon/T^4 = 0.75 \pm 0.17$ and 
$\Delta (\epsilon -3 p)/T^4 = 0.623 \pm 0.056$.
In Fig.~\ref{fig2} (right), we also show the results at $N_t=4$ obtained in Ref.~\cite{ejiri98}.
Because the data at $N_t=4$ turned out to be not far from the fitting lines in Fig.~\ref{fig2} (right), we also tried fits including the data at $N_t=4$.
The results are stable under the change of the fitting range, though the errors are not quite small yet.

\section{Gradient flow method for the latent heat}
\label{sec:flow}

\begin{figure}[t]
\begin{center}
\vspace*{-5mm}
\centerline{
\includegraphics[width=70mm,clip]{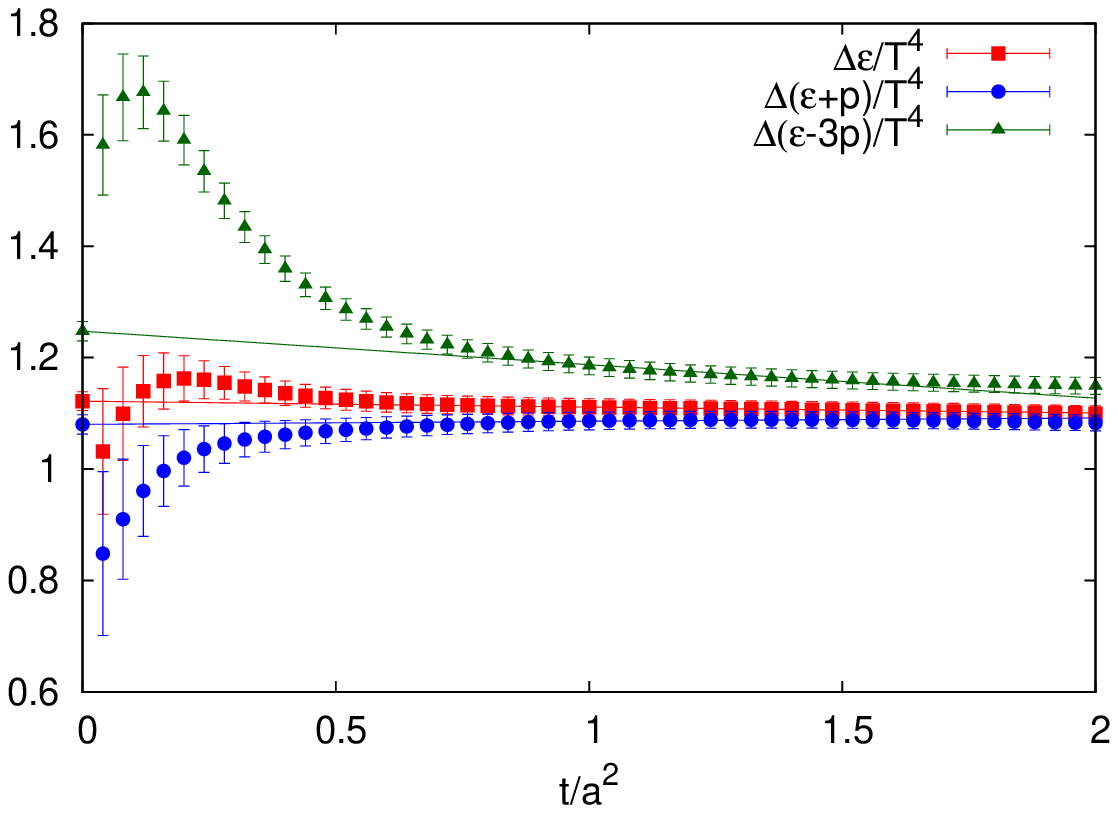}
\hspace*{-3mm}
\includegraphics[width=70mm,clip]{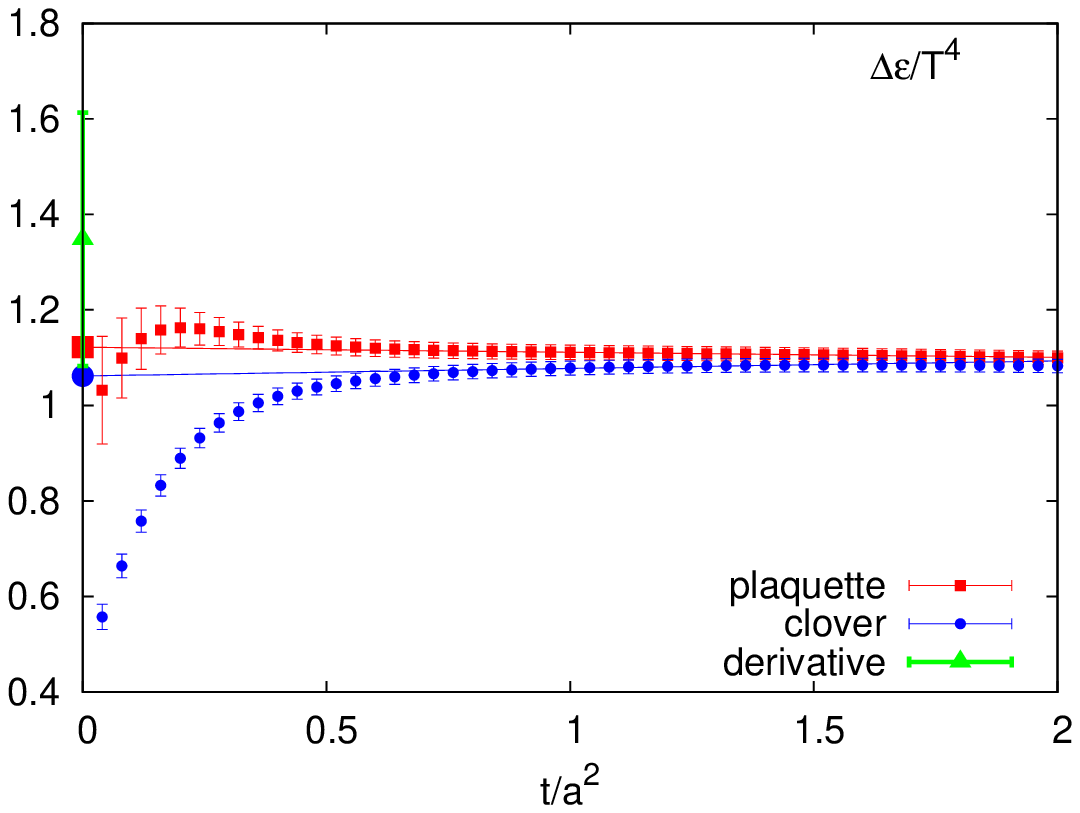}
}
\vspace{-3mm}
\caption{Left: $\Delta \epsilon/T^4$ (red), $\Delta (\epsilon + p)/T^4$ (blue) and 
$\Delta (\epsilon -3 p)/T^4$ (green) as functions of the flow time measured 
on the $96^3 \times 12$ lattice, where $G_{\mu \nu}(t,x)$ is defined by the plaquette.
Right: The latent heat $\Delta \epsilon/T^4$ constructed by the plaquette operator (red) and the clover-shaped operator (blue) on the $96^3 \times 12$ lattice. 
The triangle green symbol is the result by the derivative method obtained on the same lattice.
}
\label{fig3}
\end{center}
\end{figure}

Next, we calculate the latent heat and pressure gap by the gradient flow method \cite{suzuki13}.
The energy-momentum tensor (EMT) $T_{\mu \nu}^R$ can be constructed by the following
dimension $4$ gauge-invariant local operators,
$U_{\mu\nu}(t,x)\equiv G_{\mu\rho}(t,x)G_{\nu\rho}(t,x)
-\frac{1}{4}\delta_{\mu\nu}G_{\rho\sigma}(t,x)G_{\rho\sigma}(t,x)$
and $E(t,x)\equiv\frac{1}{4}G_{\mu\nu}(t,x)G_{\mu\nu}(t,x)$,
where $G_{\mu \nu}(t,x)$ is the field strength of flowed gauge field at the flow time $t$.
The square of $G_{\mu \nu}(t,x)$ can be defined by the plaquette or the clover operator.
We then have
\begin{eqnarray}
T_{\mu\nu}^R(x)
=\lim_{t\to0}\left\{\frac{1}{\alpha_U(t)}U_{\mu\nu}(t,x)
   +\frac{\delta_{\mu\nu}}{4\alpha_E(t)}
   \left[E(t,x)-\left\langle E(t,x)\right\rangle_0 \right]\right\},
\label{eq:(4)}
\end{eqnarray}
where the perturbative coefficients are 
$
\alpha_U(t) = \bar{g}(1/\sqrt{8t})^2
\left[1+2b_0 \bar{s}_1 \bar{g}(1/\sqrt{8t})^2+O(\bar{g}^4)\right], 
$
and
$
\alpha_E(t) = \frac{1}{2b_0}\left[1+2b_0 \bar{s}_2
\bar{g}(1/\sqrt{8t})^2+O(\bar{g}^4)\right]
$
\cite{suzuki13}.
Here, $\bar{g}(q)$ denotes the running gauge coupling in the
$\overline{\rm MS}$ scheme with the choice, $q=1/\sqrt{8t}$, and
$\bar{s}_1=\frac{7}{22}+\frac{1}{2}\gamma_E-\ln2\simeq -0.08635752993$,
$\bar{s}_2=\frac{21}{44}-\frac{b_1}{2b_0^2}=\frac{27}{484}\simeq0.05578512397$,
with 
$b_0=\frac{1}{(4\pi)^2}\frac{11}{3}N_c$,
$b_1=\frac{1}{(4\pi)^4}\frac{34}{3}N_c^2$,
and~$N_c=3$, which are the same as those used in the calculation of Ref.~\cite{flowqcd14}. 
The thermodynamic quantities
are obtained from the diagonal elements of the EMT, 
$
\epsilon = -\left\langle T_{00}^R(x)\right\rangle, 
\hspace{1mm}
p = \sum_{i=1,2,3}\langle T_{ii}^R(x)\rangle /3.
$

Separating configurations into hot and cold phases, we obtain $\Delta \epsilon/T^4$ and 
$\Delta p/T^4$.
The red, blue and green symbols in the left figure of Fig.~\ref{fig3} are the preliminary results of
$\Delta \epsilon/T^4$, $\Delta (\epsilon + p)/T^4$ and  $\Delta (\epsilon -3 p)/T^4$, 
respectively, as functions of the flow time $t$ computed on the $96^3 \times 12$ lattice, 
where $G_{\mu \nu}(t,x)$ is constructed by the plaquette operator.
As seen in Fig.~\ref{fig3} (left), the difference between $\Delta (\epsilon + p)/T^4$ and  $\Delta (\epsilon -3 p)/T^4$ 
becomes smaller as increasing $t$. 
This indicates that the pressure gap $\Delta p$ will vanish at large $t$. 
Since the lattice artifact is large when the smearing length $\sqrt{8t}$ is small 
in comparison with the lattice spacing, we perform the $t \to 0$ extrapolation omitting 
the data at small $t$. The symbols on the vertical axis are the results at $t=0$ obtained 
by fitting the data with a straight line.  
We moreover compare the results of $\Delta \epsilon/T^4$ constructed by the plaquette 
operator (red), by the clover-shaped operator (blue) and calculated by the derivative method 
(green) on the $96^3 \times 12$ lattice in Fig.~\ref{fig3} (right).
These results are consistent within the errors. 
Repeating this analysis on smaller lattices, we find that $\Delta p$ at finite $t$ decreases as $N_t$ increases. 
Hence, the continuum extrapolation will be a key point in the next step. 
The choice of the fit range in the extrapolation is also a problem we must discuss, which may be a source of the systematic error.

\section{Summary}
\label{sec:conclusion}

We calculated the latent heat and pressure gap between two phases at the first order phase transition point of SU(3) gauge theory by the derivative method with nonperturabative anisotropy coefficients.
We performed simulations around the phase transition point on lattices with $N_t=6,$ $8$ 
and $12$, and extrapolate the results to the continuum limit. 
The spatial volume dependence in the latent heat is found to be small. 
We confirmed that the pressure gap vanishes at all values of $N_t$. 
We moreover tested the gradient flow method for the calculation of the latent heat 
and compared with the results by the derivative method.
The preliminary results are consistent within the error.

\paragraph{Acknowledgments}
This work is in part supported by JSPS KAKENHI Grant 
(Nos.\ 25800148, 26287040, 26400244, 26400251, 15K05041, 16H03982).
Computations are performed at High Energy Accelerator Research Organization (KEK) %(Large Scale Simulation Program 
(Nos.\ 15/16-25, 15/16-T07).

\end{document}